# Spin-dependent transport in van der Waals magnetic tunnel junctions with Fe$_3$GeTe$_2$ electrodes

*Xinlu Li,[†] Evgeny Y. Tsymbal,[‡] Jing-Tao Lü,[†] Jia Zhang,[\*,†] Long You,[§] and Yurong Su[\*,§]*

[†]School of Physics and Wuhan National High Magnetic Field Center, Huazhong University of Science and Technology, Wuhan 430074, China

[‡]Department of Physics and Astronomy & Nebraska Center for Materials and Nanoscience, University of Nebraska, Lincoln, Nebraska 68588, USA

[§]School of Optical and Electronic Information, Huazhong University of Science and Technology, Wuhan 430074, China

ABSTRACT: Van der Waals (vdW) heterostructures, stacking different two-dimensional materials, have opened up unprecedented opportunities to explore new physics and device concepts. Especially interesting are recently discovered two-dimensional magnetic vdW materials, providing new paradigms for spintronic applications. Here, using density functional theory (DFT) calculations, we investigate the spin-dependent electronic transport across vdW magnetic tunnel junctions (MTJs) composed of Fe$_3$GeTe$_2$ ferromagnetic electrodes and a graphene or hexagonal boron nitride (*h*-BN) spacer layer. For both types of junctions, we find that the junction resistance changes by thousands of percent when the magnetization of the electrodes is switched from parallel to antiparallel. Such a giant tunneling magnetoresistance (TMR) effect is driven by dissimilar electronic structure of the two spin-conducting channels in Fe$_3$GeTe$_2$, resulting in a mismatch between the incoming and outgoing Bloch states in the electrodes and thus suppressed transmission for an antiparallel-aligned MTJ. The vdW bounding between electrodes and a spacer layer makes this result virtually independent of the type of the spacer layer, making the predicted giant TMR effect robust with respect to strain, lattice mismatch, interface distance and other parameters which may vary in the experiment. We hope that our results will further stimulate experimental studies of vdW MTJs and pave the way for their applications in spintronics.

The emergence of magnetic two-dimensional (2D) van der Waals (vdW) materials offers exciting opportunities for exploring new physical phenomena and potential applications.[1,2] Among these materials are $Cr_2Ge_2Te_6$,[3] $VSe_2$[4] and $CrI_3$,[5,6] where a long-range 2D magnetic order has recently been discovered. Heterostructures based on these vdW materials have revealed interesting functional properties.[7,8] In particular, 2D magnetic layers have been employed as electrodes or barriers in vdW magnetic tunnel junctions (MTJs), expanding the field of spin-dependent transport in MTJs beyond conventional transition metal ferromagnetic electrodes and oxide barrier layers. For instance, using $CrI_3$ as a barrier layer in graphite/$CrI_3$/graphite tunnel junctions has been reported to produce a huge tunneling magnetoresistance (TMR) over thousands of percent at low temperature.[9,10] In these tunnel junctions, $CrI_3$ served as a spin-filter and TMR was associated with a change of the $CrI_3$ magnetic ordering from antiferromagnetic to ferromagnetic under the influence of an applied magnetic field.[11] This is different from the "conventional" TMR effect in MTJs where two ferromagnetic electrodes are realigned by an applied magnetic field, resulting in a change of tunneling resistance,[12] and the related effects in ferroelectric tunnel junctions with ferromagnetic electrodes.[13] Very recently, TMR of 160% has been experimentally observed in $Fe_3GeTe_2|h\text{-}BN|Fe_3GeTe_2$ MTJs at low temperature.[14] The observed effect resembles the conventional TMR associated with the change of magnetic ordering of ferromagnetic electrodes. However, essentially the spin-dependent transport mechanism in these vdW MTJs has not been elucidated. There are a number of fundamental questions, which need to be addressed in order to guide further developments in the field of vdW MTJs. Among these questions are the effect of the spin-dependent electronic structure of $Fe_3GeTe_2$ on TMR, the importance of the symmetry selection rules which are known to control TMR in MTJs with transition metal electrodes and crystalline tunnel barriers,[15-19] the role of the tunnel barrier layer and its electronic structure,[15,19] and the effect of the interface bonding on magnetoresistive properties.[16]

Answering these questions are especially important from the practical perspective. Among known 2D ferromagnetic vdW materials, $Fe_3GeTe_2$ exhibits relatively high Curie temperature around 220 K in its bulk state.[20-22] Moreover, it has been shown that the Curie temperature of $Fe_3GeTe_2$ can be raised up to room temperature by ionic gating.[23] The metallic nature of $Fe_3GeTe_2$ enables using this material as a magnetic electrode in vdW MTJs, which has advantages over insulating $CrI_3$ used as a spin-filter barrier. First, in $CrI_3$-based MTJs, a large magnetic field (~ 1T) is required to switch the antiferromagnetic ground state to ferromagnetic.[9,10] Second, $CrI_3$-based MTJs are volatile, i.e. the magnetic field needs to be maintained to preserve the ferromagnetic order, while $Fe_3GeTe_2$-based MTJs are non-volatile due two stable magnetization configurations (parallel and antiparallel) in the absence of applied field.

Driven by these motivations, in this letter, we investigate the spin-dependent transport across vdW MTJs with $Fe_3GeTe_2$ electrodes and two representative non-magnetic 2D materials graphene (Gr) and *h*-BN as spacer layers. We demonstrate the emergence of a giant TMR in these junctions which is controlled by the electronic structure of $Fe_3GeTe_2$ and largely independent of the nature of the spacer layer. These results open interesting perspectives for further experimental exploration of MTJs based on 2D magnetic vdW materials.

First-principles calculations are performed using the Quantum ESPRESSO package[24] with PBE-GGA exchange correlation potential[25] and ultrasoft pseudopotential[26]. The electronic structure of bulk $Fe_3GeTe_2$ is self-consistently calculated with the lattice parameters being fixed at their experimental values of $a = b = 3.991$Å and $c = 16.33$ Å.[20] The Monkhorst k-point mesh for the self-consistent calculation is $16\times16\times4$ and the plane-wave cutoff is 40 Ry.

In a MTJ, the $\sqrt{3} \times \sqrt{3}$ in-plane unit cell of graphene and h-BN is matched at the interface with that of the $Fe_3GeTe_2$ electrode. The interface spacing between $Fe_3GeTe_2$ and graphene or h-BN is relaxed in the presence of the vdW interaction. The electronic structure of $Fe_3GeTe_2|Gr|Fe_3GeTe_2$ and $Fe_3GeTe_2|h$-$BN|Fe_3GeTe_2$ supercells is self-consistently calculated using a fine k-point mesh of 16×16×1. Then, the electron transmission is obtained using the wave-function scattering method,[27] by matching the wave functions between left and right $Fe_3GeTe_2$ electrodes and the scattering region, representing MTJs. The two-dimensional Brillouin zone (2DBZ) is sampled using a uniform mesh of 200×200 k-points. The spin-dependent ballistic conductance of the MTJs is obtained by the summation of transmission over the 2DBZ:

$$G_\sigma = \frac{e^2}{h} \sum_{\mathbf{k}_\parallel} T_\sigma(\mathbf{k}_\parallel)$$

where $T_\sigma(\mathbf{k}_\parallel)$ is the spin and k-resolved transmission probability for an electron at the Fermi energy with spin σ and Bloch wave vector $\mathbf{k}_\parallel = (k_x, k_y)$, $e$ is the elementary charge, and $h$ is the Plank constant. The TMR ratio is defined as $TMR=(G_P - G_{AP})/G_{AP}$, where $G_{AP}$ and $G_P$ are conductances for parallel and antiparallel alignment of magnetization of the $Fe_3GeTe_2$ electrodes, respectively.

The crystal structure of bulk $Fe_3GeTe_2$ is depicted in Figure 1a. $Fe_3GeTe_2$ has a layered hexagonal crystal structure which contains $Fe_3Ge$ slabs separated by the vdW bonded Te layers. The Fe atoms in the unit cell are located in two inequivalent

Wyckoff sites, denoted in Figure 1a as Fe1 and Fe2, respectively. The Fe1-Fe1 pair bonds cross a hexagonal network built by covalently bonded Fe2 and Ge atoms. The calculated band structure and density of states of bulk $Fe_3GeTe_2$ are shown in Figure 1b and 1c, respectively. The presence of a sizable electron density at the Fermi energy and the exchange splitting of the spin bands indicate that $Fe_3GeTe_2$ is a ferromagnetic metal. The calculated band structure agrees well with the previous calculation results and the experimental band structures obtained by ARPES.[28]

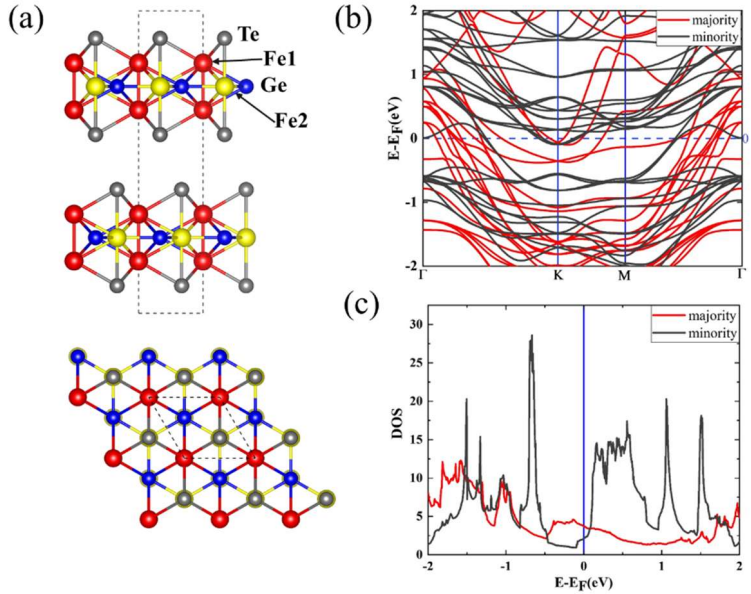

**Figure 1.** (a) Side and top views of the atomic lattice of $Fe_3GeTe_2$. The dashed rectangular and rhombic shapes denote the crystal unit cell. (b) Majority-spin (black curves) and minority-spin (red curves) band structure of $Fe_3GeTe_2$. (c) Spin resolved density of states of bulk $Fe_3GeTe_2$.

In a MTJ, transmission and magnetoresistance at zero bias are affected by the spin-polarized Fermi surface of the electrodes. Figure 2a and 2b show the three-dimensional Fermi surface of bulk $Fe_3GeTe_2$ for majority- and minority-spin electrons, respectively. It is seen that while the majority-spin Fermi surface (Figure 2a) consists of many Fermi sheets and covers a large portion of the 2DBZ, the minority-spin Fermi surface (Figure 2b) consists of isolated sheets localized in a small portion of the 2DBZ. These bands represent the incoming and outgoing Bloch states, which characterize transport across a MTJ. The number of available Bloch states (conduction channels) at each $k_\parallel$ point can be obtained by calculating ballistic transmission in bulk $Fe_3GeTe_2$. Figure 2c and 2d show the $k_\parallel$-resolved ballistic transmission of bulk $Fe_3GeTe_2$ for majority- and minority-spin channels, respectively. It is seen that the transmission at each $k_\parallel$ point is an integer, mirroring the corresponding spin-resolved Fermi surfaces shown in Figure 2a,b.

By comparing the distribution of the conduction channels in $Fe_3GeTe_2$ for majority- and minority-spin electrons over the 2DBZ (Figure 2c and 2d, respectively), we can make qualitative conclusions about TMR in MTJs based on these electrodes, even without knowing the explicit electronic structure of the whole tunnel junction and calculating transmission across it. Due to spin and $k_\parallel$ being conserved in the transmission process, transmission between parallel-aligned ferromagnetic electrodes is expected to be much larger than transmission between antiparallel-aligned electrodes. This is due to the fact that in the latter case a mismatch between the incoming and outgoing Bloch sates for majority-spin (Figure 2c) and minority-spin (Figure 2d) channels would strongly reduce the transmission. As seen from Figure 2c and 2d, while the majority-spin channel has multiple bands at the Fermi energy covering the large portion of the 2DBZ, the minority-spin channel has only a few states available, resulting in a large area of the 2DBZ with no overlap. Based on this observation, one can expect high magnetoresistance in 2D vdW MTJs with $Fe_3GeTe_2$ electrodes.

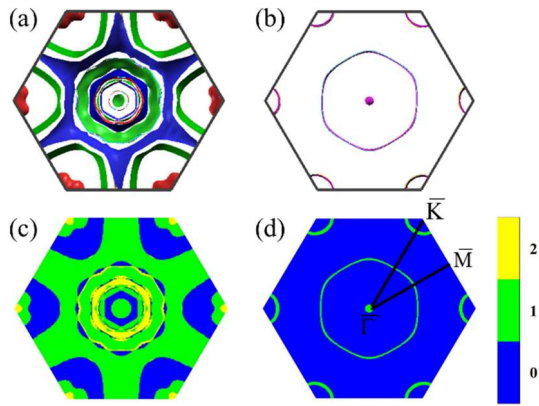

**Figure 2.** Majority-spin (a) and minority-spin (b) Fermi surfaces of $Fe_3GeTe_2$ (plotted using the xcrysden package[29]). Different sheets of the Fermi surface, which belong to different bands, are indicated in color. Ballistic transmission for majority-spin (c) and minority-spin (d) channels in bulk $Fe_3GeTe_2$, representing the number of conducting channels in 2DBZ indicated in different color.

Next, we explore MTJs based on $Fe_3GeTe_2$ electrodes and graphene or *h*-BN barrier layers. The structures of MTJs are shown in Figure S1 in the Supporting Information. Matching of the in-plane lattice leads to graphene and *h*-BN being compressed by 6.8% and 9.3%, respectively. These large values of the compressive strain are unlikely to occur in the experimental conditions, where the lattices of the electrodes and the barrier would be mismatched due to a weak vdW type bonding, but are required to maintain periodic boundary conditions in our computations. We have compared the band structure of graphene before and after the interface matching and found that there is no significant change in its electronic structure as shown in Figure S2 in the Supporting Information. For *h*-BN, a slightly larger interface mismatch leads to some changes in the electronic structure but the insulating nature of this material remains preserved as it is shown in Figure S3 in the Supporting Information. The optimized interface distance between the $Fe_3GeTe_2$ electrode and the spacer layer is about 3.50 Å, revealing a weak interface bonding across the interface. For comparison, in a Fe|MgO|Fe MTJ, the Fe-MgO interface distance is about 2.16 Å.[15] The weak interface interaction between the electrodes and the spacer layer in vdW MTJs makes the electronic structure of the junctions being simply composed of the electronic structures of its constituents. This behavior is evident from the weight-projected band structure of the $Fe_3GeTe_2$|Gr|$Fe_3GeTe_2$ MTJ shown in Figure 3. There are no obvious effects of hybridization and band offset between $Fe_3GeTe_2$ and graphene. The graphene bands in the MTJ remain mostly unchanged compared to the isolated graphene. This provides evidence of the rather weak interface interaction between $Fe_3GeTe_2$ and the spacer layer. We note that while the Dirac cone of graphene lies at the $\bar{\text{K}}$ point for the original 1×1 cell, it is located at the $\bar{\Gamma}$ point for the $\sqrt{3}\times\sqrt{3}$ unit cell when it is matched to the unit cell of Fe$_3$GeTe$_2$ (Figure S2).

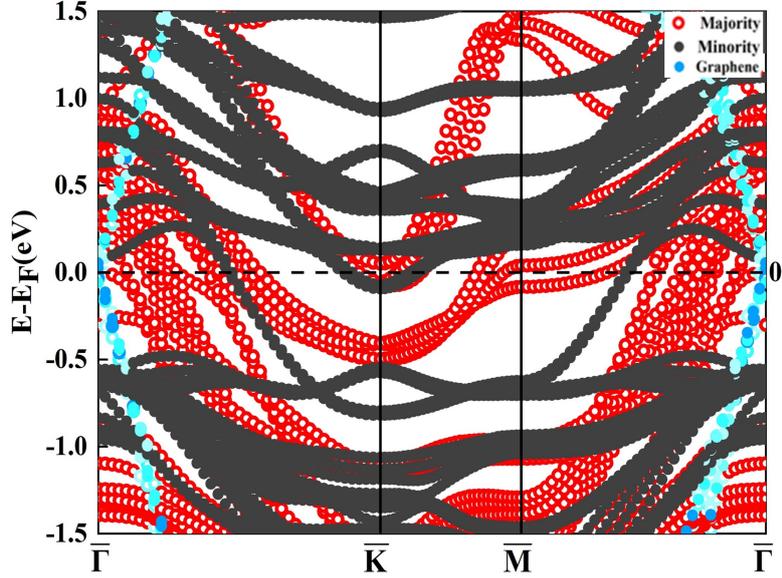

**Figure 3.** The weight projected band structure in the magnetic junction of Fe$_3$GeTe$_2$|Gr|Fe$_3$GeTe$_2$. The red circles and solid black circles represent the majority and minority band structures of Fe$_3$GeTe$_2$, respectively. The blue dots demonstrate the projected band weight on graphene. Energy zero indicates the position of Fermi energy.

Figure 4a-c show the calculated spin- and $\mathbf{k}_{\parallel}$-resolved electron transmission across a Fe$_3$GeTe$_2$|Gr|Fe$_3$GeTe$_2$ MTJ for parallel (Figure 4a,b) and antiparallel (Figure 4c) magnetization of the electrodes. It is seen that both majority- and minority-spin transmissions resemble the corresponding distribution of the conducting channels in the Fe$_3$GeTe$_2$ electrodes (compare Figure 4a to 2c and Figure 4b to 2d). Although the graphene barrier layer filters out electronic transmission at the periphery of the 2DBZ, overall, it does qualitatively change the balance between the majority- and minority-spin contributions. Consistent with our discussion above, the antiparallel transmission (Figure 4c) is strongly reduced and qualitatively resembles the minority-spin transmission (Figure 4b).

We note that although graphene is a conductor, transmission across a Fe$_3$GeTe$_2$|Gr|Fe$_3$GeTe$_2$ junction is dominated by quantum-mechanical tunneling. The Fermi surface of graphene is reduced to a single $\bar{\Gamma}$ point, and there are no other available propagating states in graphene in the entire 2DBZ. Therefore, for $\mathbf{k}_\parallel \neq 0$, the transmission across the graphene layer is controlled by its evanescent states, in which decay rate is determined by a barrier height. The latter is increasing away from the $\bar{\Gamma}$ point, resulting in the reduced transmission at the periphery of the 2DBZ.

Figure 4d-f show the calculated $\mathbf{k}_\parallel$-resolved transmission across a Fe$_3$GeTe$_2$|h-BN|Fe$_3$GeTe$_2$ MTJ. Contrary to graphene, h-BN is an insulator, however, the transmission patterns for the two junctions with graphene and h-BN barrier layers are very similar. Although h-BN does not have propagating states at the Fermi energy, similar to graphene, as it is shown in Figure S3, h-BN has a lower barrier at the center of the 2DBZ ($\bar{\Gamma}$ point), resulting in the enhanced transmission around this area. These results indicate that the key features of spin-dependent tunneling in the vdW MTJs with ferromagnetic Fe$_3$GeTe$_2$ electrodes are largely independent of the nature of the barrier. This statement is further confirmed by replacing graphene or h-BN with a vacuum barrier layer. The resulting $\mathbf{k}_\parallel$-resolved transmission for a Fe$_3$GeTe$_2$|Vacuum|Fe$_3$GeTe$_2$ MTJ is shown in Figure S4 of Supporting Information, which reveals a similar transmission pattern as in the case of Fe$_3$GeTe$_2$|Gr|Fe$_3$GeTe$_2$ and Fe$_3$GeTe$_2$|h-BN|Fe$_3$GeTe$_2$ MTJs.

The total spin-resolved transmission is calculated and summarized in Table 1. It is seen that for both MTJs, the total transmission for the parallel magnetization state is two orders of magnitude larger than that for the antiparallel state. The resulting TMR ratios for MTJs with graphene and h-BN spacer layers are 3621% and 6256%, respectively. These results suggest that in the 2D vdW MTJs, TMR is very large regardless of the spacer layer. This key feature is different from the conventional MgO-based MTJs, in which the high TMR ratio originates from the spin-polarized transmission of the $\Delta_1$-symmetry band across an MgO tunnel barrier.[15] In addition, in traditional MTJs such as Fe|MgO|Fe, the interface effects have crucial importance for

the spin-dependent transport and the resulting TMR is affected by the interface resonant states.[16,30] However, in MTJs with 2D ferromagnetic electrodes, the interface effects are expected to be suppressed due to a weak vdW interaction. As a result, the transport properties are mainly determined by the spin-dependent electronic properties of bulk $Fe_3GeTe_2$.

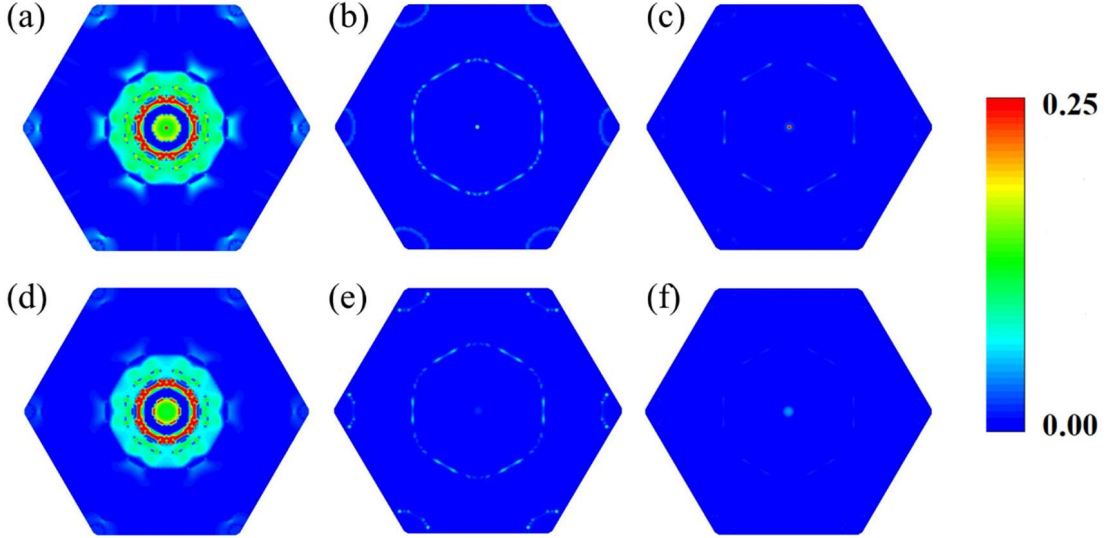

**Figure 4.** The $\mathbf{k}_\parallel$-resolved electron transmission distribution in the 2DBZ for $Fe_3GeTe_2|Gr|Fe_3GeTe_2$ (a-c) and $Fe_3GeTe_2|h\text{-}BN|Fe_3GeTe_2$ (d-f) MTJs. (a), (d): majority-spin transmission for parallel magnetization state; (b), (e): minority-spin transmission for parallel magnetization state; (c) and (f) either-spin transmission for antiparallel magnetization state.

**Table 1.** The calculated electron transmission across $Fe_3GeTe_2|Gr|Fe_3GeTe_2$ and $Fe_3GeTe_2|h\text{-}BN|Fe_3GeTe_2$ MTJs for parallel (P) and antiparallel (AP) magnetization of $Fe_3GeTe_2$ electrodes.

| MTJ structures | P transmission | AP transmission | TMR (%) |
| --- | --- | --- | --- |
| $Fe_3GeTe_2|Gr|Fe_3GeTe_2$ | $2.60\times10^{-2}$ | $6.98\times10^{-4}$ | 3621 |
| $Fe_3GeTe_2|h\text{-}BN|Fe_3GeTe_2$ | $2.15\times10^{-2}$ | $3.38\times10^{-4}$ | 6256 |

Previous studies of MgO-based MTJs have shown that TMR may change sizably depending on the energy position of the Fermi level. Moreover, in practice, the Fermi level of the junction may be different from the ideal case due to structural imperfections, doping, etc. In order to clarify the effect of the Fermi energy and prove that our conclusions are robust with respect to this characteristic, we calculate the transmission as a function of energy. The transmission and TMR of the $Fe_3GeTe_2$|Gr|$Fe_3GeTe_2$ and $Fe_3GeTe_2$|$h$-BN|$Fe_3GeTe_2$ MTJs are calculated within the energy window ranging from -0.3eV to +0.3eV (The original Fermi level, $E_F$, is set at zero). The calculated transmission and TMR ratio as a function of energy are plotted in Figure 5 and are listed in Tables S1 and S2 in Supporting Information. We see that the transmission as a function of energy shows similar trends for the two junctions. In both cases, the TMR value is reduced at positive energies but dramatically enhanced at negative energies. In the energy range below $E_F$-0.2eV (between $E_F$-0.2eV and $E_F$-0.3 eV in Figure 5), the TMR ratio becomes virtually infinite. This is due to the fact that the antiparallel-state transmission in this energy range becomes zero as is seen in Figure 5. The similar trend is demonstrated by a $Fe_3GeTe_2$|Vacuum|$Fe_3GeTe_2$ MTJ, where the TMR ratio goes to infinity at negative energies below $E_F$-0.2eV (Figure S5).

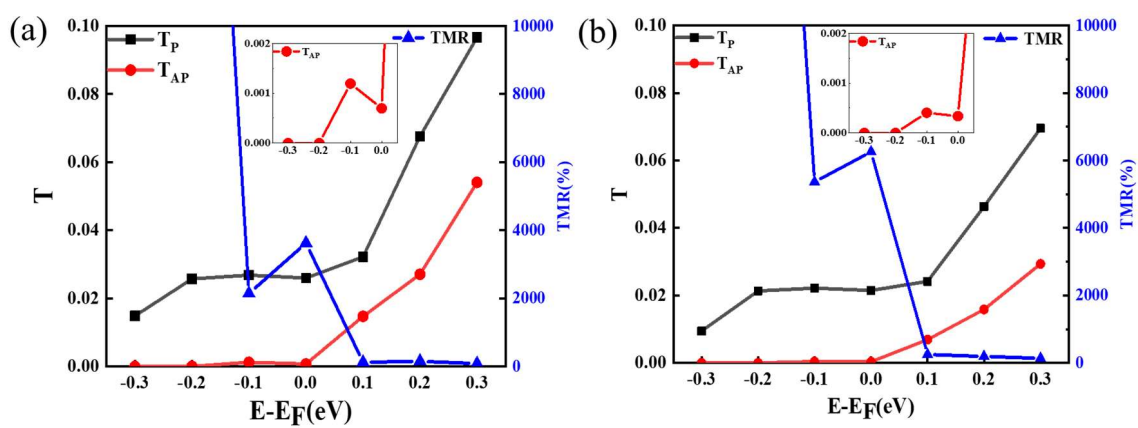

**Figure 5.** The electron transmission for the P (black squares) and AP (red circles) magnetization states and the corresponding TMR ratio (blue triangles, refer to the right axis) as a function of energy for $Fe_3GeTe_2$|Gr|$Fe_3GeTe_2$ (a) and $Fe_3GeTe_2$|$h$-BN|$Fe_3GeTe_2$ (b) MTJs.

The insets show the details of the AP state transmission in the energy range of $E_F$-0.3eV to $E_F$.

A complete mismatch between the Fermi surfaces of the two spin channels is the reason for zero transmission for the antiparallel magnetization. A similar behavior is expected to occur in MTJs with half-metallic ferromagnets, such as Heusler alloys $Co_2MnSi$, $Co_2MnGe$ *etc.*, which have only one spin band at the Fermi energy. However, in MTJs with half-metallic Heusler electrodes, e.g., $Co_2MnSi|MgO|Co_2MnSi$, interface states diminish half-metallicity of the Heusler alloy through the interface bonding.[31] Thanks to the weak interface interaction in 2D vdW MTJs, the interface bonding effects are suppressed and the transmission is entirely controlled by the bulk properties of ferromagnetic electrodes. Figure 6 shows that calculated Fermi surface of $Fe_3GeTe_2$ at different energies. It is seen that in the energy range of $E_F$-0.2 and $E_F$-0.3 eV, the majority- and minority-spin conduction channels do not overlap in the 2DBZ. As a result, there are no electrons that can successfully pass through the MTJ when the magnetization of the $Fe_3GeTe_2$ electrodes is antiparallel, which leads to the infinite TMR in the corresponding energy range.

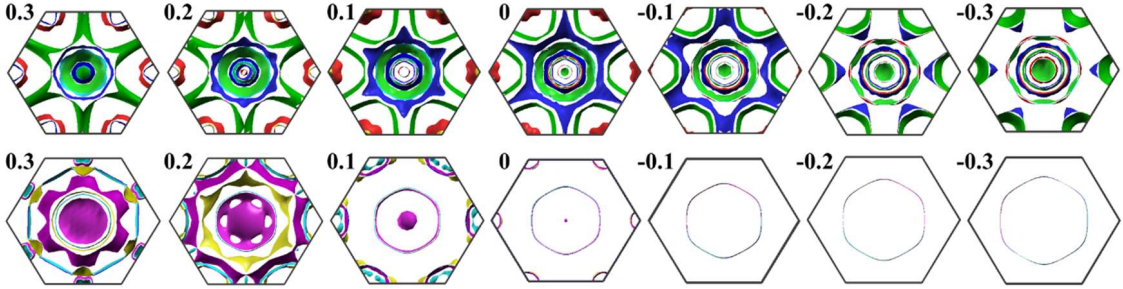

**Figure 6.** The majority-spin (top panel) and minority-spin (bottom panel) Fermi surfaces of $Fe_3GeTe_2$ at different energies ranging from $E_F$-0.3 to $E_F$+0.3 eV, where "0" indicates the Fermi energy. Colors indicate the Fermi surfaces belonging to different bands.

Using the calculated transmission as a function of energy, it is straightforward to calculate the *I-V* curves. Under a small bias voltage, ignoring the non-equilibrium effect, the electric current per unit cell under bias voltage *V* is given by

$$I_{P,AP} = \frac{e}{h} \int_{E_F - \frac{eV}{2}}^{E_F + \frac{eV}{2}} T_{P,AP}(E) dE .$$

Here $T_P(E)$ ($T_{AP}(E)$) is the transmission as a function of energy $E$ for the P (AP) state of MTJ (Figure 5). And the current On/Off ratio can be defined to be the ratio between current for P and AP state: $I_P/I_{AP}$. The *I-V* curves for P and AP states for two magnetic junctions are shown in Figure S6 in Supporting Information. We can see that in a wide range of bias voltage, the On/Off ratio is over 10 and the half-width when On/Off ratio drops to half of its maximum value is around 0.04 V for both MTJs.

In summary, the spin-dependent transport properties in 2D vdW magnetic tunnel junctions with $Fe_3GeTe_2$ electrodes have been investigated using first-principles calculations. It is predicted that regardless of the spacer layer, $Fe_3GeTe_2|Gr|Fe_3GeTe_2$ and $Fe_3GeTe_2|h\text{-}BN|Fe_3GeTe_2$ MTJs exhibit the TMR ratio as high as thousands of percent, which is the consequence of the significantly different Fermi surfaces of the two spin-conduction channels. The key mechanism responsible for the giant TMR effect in the studied vdW MTJs is solely relies on the spin-dependent electronic structure of ferromagnetic $Fe_3GeTe_2$ electrodes. It is also found that in a certain energy range, the complete mismatch of the majority- and minority-spin Fermi surfaces results in zero transmission for the antiparallel magnetization state and therefore leads to the infinite TMR. Our results provide an important insight for further experimental investigations of 2D vdW MTJs which may lead useful applications in spintronics.

## ■ ASSOCIATED CONTENT

**Supporting Information**

Magnetic junction structures with graphene, *h*-BN and vacuum sandwiched between two $Fe_3GeTe_2$ electrodes; the band structure of $\sqrt{3} \times \sqrt{3}$ graphene before and after it is compressed by 6.8%; the band structure of $\sqrt{3} \times \sqrt{3}$ *h*-BN before and after it is compressed by 9.3%; spin-dependent transport properties in $Fe_3GeTe_2|Vacuum|Fe_3GeTe_2$ magnetic junction; transmission data as a function of energy in $Fe_3GeTe_2|Gr|Fe_3GeTe_2$ and $Fe_3GeTe_2|h\text{-}BN|Fe_3GeTe_2$ magnetic junctions; I-V curves in $Fe_3GeTe_2|Gr|Fe_3GeTe_2$ and $Fe_3GeTe_2|h\text{-}BN|Fe_3GeTe_2$ magnetic junctions (PDF)

## ■ AUTHOR INFORMATION

**Corresponding Authors**

*(J. Z.) E-mail: jiazhang@hust.edu.cn

*(Y. S.) E-mail: suyr@hust.edu.cn

**Notes**

The authors declare no competing financial interest.

## ■ ACKNOWLEDGMENTS

J. Zhang is supported by the National Natural Science Foundation of China with grant No. 11704135. The calculations in this work are partly performed at National Supercomputer Center in Tianjin, TianHe-1(A) China.

# Supporting Information for

# Spin-dependent transport in van der Waals magnetic junctions with Fe$_3$GeTe$_2$ electrode

*Xinlu Li,[†] Evgeny Y. Tsymbal,[‡] Jing-Tao Lü,[†] Jia Zhang,[*,†] Long You,[§] and Yurong Su[*,§]*

[†]School of Physics and Wuhan National High Magnetic Field Center, Huazhong University of Science and Technology, Wuhan 430074, China

[‡]Department of Physics and Astronomy & Nebraska Center for Materials and Nanoscience, University of Nebraska, Lincoln, Nebraska 68588, USA

[§]School of Optical and Electronic Information, Huazhong University of Science and Technology, Wuhan 430074, China

**Corresponding Authors**

*(J. Z.) E-mail: jiazhang@hust.edu.cn.

*(Y. S.) E-mail: suyr@hust.edu.cn.

## 1. The illustration of magnetic junction structures with graphene, *h*-BN and vacuum sandwiched between two Fe$_3$GeTe$_2$ electrodes.

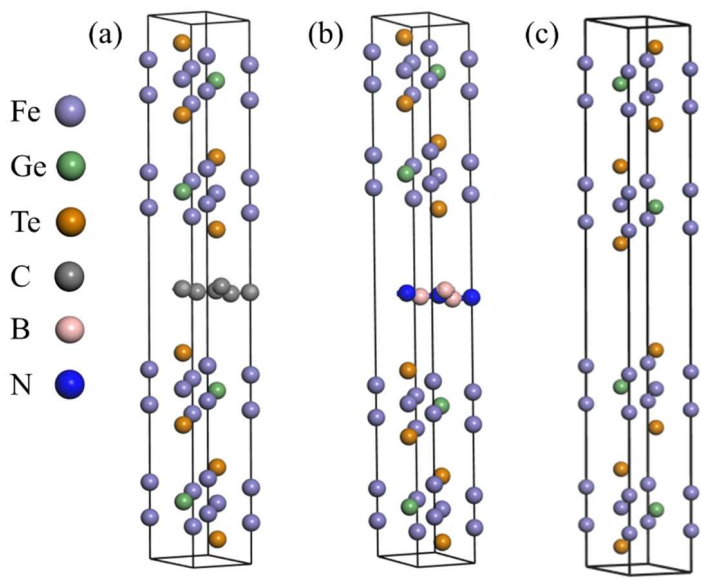

Figure S1. The magnetic junction structures of (a) Fe$_3$GeTe$_2$|Graphene|Fe$_3$GeTe$_2$, (b) Fe$_3$GeTe$_2$|*h*-BN|Fe$_3$GeTe$_2$ and (c) Fe$_3$GeTe$_2$|Vacuum|Fe$_3$GeTe$_2$.

## 2. The band structure of $\sqrt{3} \times \sqrt{3}$ graphene before and after it is compressed by 6.8% to match the Fe₃GeTe₂ interface. The band structure of original 1×1 graphene is also plotted for comparison.

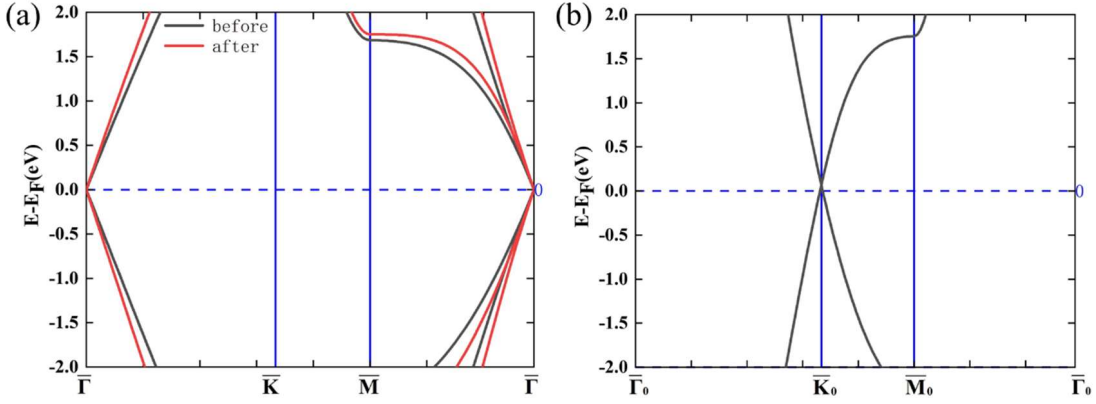

Figure S2. (a) The band structures of the $\sqrt{3} \times \sqrt{3}$ Graphene before (in black) and after (in read) it is compressed by 6.8% for the interface matching. (b) The band structures of the 1×1 graphene. The horizontal blue dash lines located at zero energy indicate the position of Fermi energy.

## 3. The band structure of $\sqrt{3} \times \sqrt{3}$ h-BN before and after it is compressed by 9.3% to match the Fe₃GeTe₂ interface. The projected band structure of the junction Fe₃GeTe₂|h-BN|Fe₃GeTe₂ is also plotted.

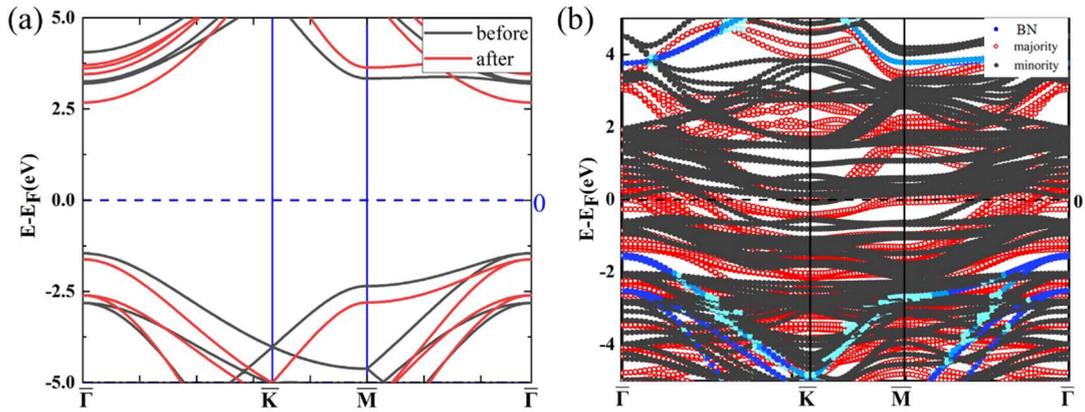

Figure S3. (a) The band structures of the $\sqrt{3} \times \sqrt{3}$ h-BN before (in black) and after (in red) it is compressed by 9.3% for the interface matching. (b) The weight projected band structure in the magnetic junction of Fe₃GeTe₂|h-BN|Fe₃GeTe₂. The red circles and the black dots represent the majority and minority band structures of the magnetic junction, respectively. The blue dots demonstrate the projected band weight of h-BN. The horizontal dash lines at zero energy indicate the position of Fermi energy which lies within the bandgap of *h*-BN.

## 4. The spin-dependent transport properties in Fe₃GeTe₂|Vacuum|Fe₃GeTe₂ magnetic junction.

In order to further demonstrate the decisive role of the Fe₃GeTe₂ electrode in the tunnel magnetoresistance (TMR) effect, as a comparison, the transmission in a magnetic junction with vacuum barrier has also been investigated (Figure S4).

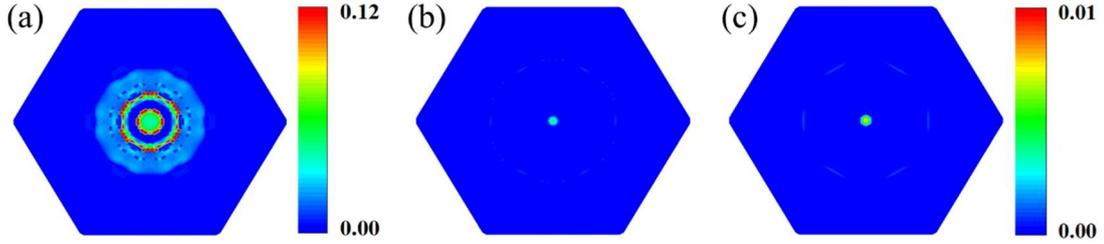

Figure S4. The $\mathbf{k}_\parallel$-resolved electron transmission distribution in 2DBZ of Fe₃GeTe₂|Vacuum|Fe₃GeTe₂ junction. (a) and (b) are the majority and minority electron transmission for the P state. (c) is the majority to minority (or minority to majority) transmission for AP state. The color bars at right side of (a) and (c) indicate the transmission intensity.

The total transmission in Fe₃GeTe₂|Vacuum|Fe₃GeTe₂ as a function of energy is calculated and plotted in Figure S5. The transmission of the junction with vacuum barrier is lower than that in MTJs with graphene or h-BN barrier due to the larger energy barrier height (at the order of several eV), but qualitatively similar energy dependence can be found. This confirms the large TMR is regardless of the specific spacer. In addition, zero transmission of AP state below $E_F$-0.2 eV results in infinite TMR, which is similar as the behavior of MTJs with graphene or *h*-BN barrier.

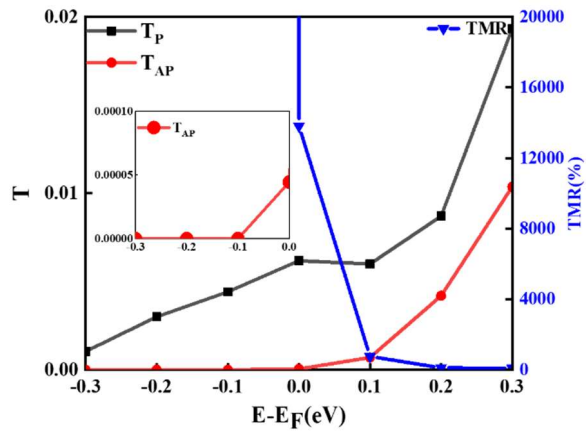

Figure S5. The electron transmission for the parallel (P) and antiparallel (AP) states and the corresponding TMR as a function of energy in the $Fe_3GeTe_2|Vacuum|Fe_3GeTe_2$ magnetic junction. The black squares and the red circles represent the transmission of P and AP state, respectively. The blue triangles represent the magnetoresistance (refer to right axis). The inset shows the details of the transmission of AP state in the range of $E_F$-0.3 eV to $E_F$.

## 5. The calculated transmission data as a function of energy in $Fe_3GeTe_2|Graphene|Fe_3GeTe_2$ and $Fe_3GeTe_2|h\text{-}BN|Fe_3GeTe_2$ magnetic junctions.

**Table S1**. The electron transmission versus energy and the corresponding TMR in $Fe_3GeTe_2|Graphene|Fe_3GeTe_2$ magnetic junction. The zero indicates the Fermi energy.

| Energy (eV) | P transmission | AP transmission | TMR (%) |
|---|---|---|---|
| +0.3 | $9.67 \times 10^{-2}$ | $5.41 \times 10^{-2}$ | 78.8 |
| +0.2 | $6.76 \times 10^{-2}$ | $2.71 \times 10^{-2}$ | 149.8 |
| +0.1 | $3.22 \times 10^{-2}$ | $1.47 \times 10^{-2}$ | 119.2 |
| 0 | $2.60 \times 10^{-2}$ | $6.98 \times 10^{-3}$ | 3621.5 |
| -0.1 | $2.68 \times 10^{-2}$ | $1.20 \times 10^{-3}$ | 2141.5 |
| -0.2 | $2.57 \times 10^{-2}$ | 0.0 | ∞ |
| -0.3 | $1.48 \times 10^{-2}$ | 0.0 | ∞ |

**Table S2**. The electron transmission versus energy and the corresponding TMR in $Fe_3GeTe_2|h\text{-}BN|Fe_3GeTe_2$ magnetic junction. The zero indicates the Fermi energy.

| Energy (eV) | P transmission | AP transmission | TMR (%) |
|---|---|---|---|
| +0.3 | $6.95 \times 10^{-2}$ | $2.93 \times 10^{-2}$ | 137.0 |
| +0.2 | $4.64 \times 10^{-2}$ | $1.58 \times 10^{-2}$ | 192.5 |
| +0.1 | $2.42 \times 10^{-2}$ | $6.91 \times 10^{-3}$ | 249.7 |
| 0 | $2.15 \times 10^{-2}$ | $3.38 \times 10^{-4}$ | 6256.5 |
| -0.1 | $2.22 \times 10^{-2}$ | $4.06 \times 10^{-4}$ | 5360.9 |
| -0.2 | $2.13 \times 10^{-2}$ | 0 | ∞ |
| -0.3 | $9.44 \times 10^{-3}$ | 0 | ∞ |

## 6. The simplified calculation of I-V curves in Fe$_3$GeTe$_2$|Graphene|Fe$_3$GeTe$_2$ and Fe$_3$GeTe$_2$|h-BN|Fe$_3$GeTe$_2$ magnetic junctions.

In order to obtain the *I-V* curve, the transmission has been linearly interpolated from the data shown in Figure 5 in the main text. By ignoring the non-equilibrium effect, the current under bias voltage *V* can be written as:

$$I_P = \frac{e}{h} \int_{E_F - \frac{eV}{2}}^{E_F + \frac{eV}{2}} T_P(E) dE$$

The On/Off ratio can be defined to be the ratio between current for P and AP state $I_P/I_{AP}$. When the On/Off ratio drops to half of maximum value, the half-width is around 0.04 *V* for both magnetic junctions.

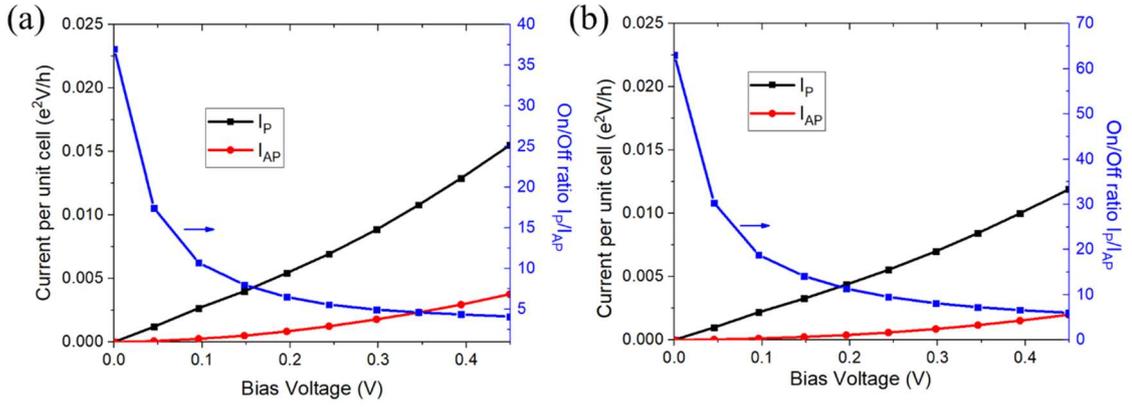

Figure S6. The calculated current and On/Off ratio as a function of bias voltage for (a) Fe$_3$GeTe$_2$|Graphene|Fe$_3$GeTe$_2$ and (b) Fe$_3$GeTe$_2$|h-BN|Fe$_3$GeTe$_2$ magnetic junctions.